\documentclass[epsf,11pt]{article}
\usepackage{epsfig}
\usepackage{amsmath}
\usepackage{amssymb}

\begin{document}

\title{The study of SU(3) super Yang-Mills quantum mechanics.
}
\author{Maciej Trzetrzelewski \footnote{trzetrzelewski@th.if.uj.edu.pl} \\
M. Smoluchowski Institute of Physics, Jagiellonian University \\
Reymonta 4, 30-059 Cracow, Poland \\
  }
  
\maketitle
\abstract{
 We present the hamiltonian study of super Yang-Mills quantum mechanics
(SYMQM). The recently introduced method based on Fock space representation
allows to analyze SYMQM numerically. The detailed analysis for SYMQM in two
dimensions for $SU(3)$ group is given.}

\section{Introduction}
Supersymmetric Yang-Mills quantum mechanics are very interesting
models since they emerge in different areas of physics. The
general, not necessarily gauge,  supersymmetric quantum mechanics
have been studied first as a laboratory of supersymmetry [1]
where in particular the exact solution for $D=2$, $SU(2)$ case was
given. By definition SYMQM are $\mathcal{N}=1$ super Yang-Mills
field quantum theories reduced from $D=d+1$ to $D=0+1$ dimensions.
Supersymmetry requires the space-time dimension to be $D=2,4,6,10$
with $\mathcal{N}=2,4,8,16$ supercharges in the resulting quantum
mechanics respectively. The rotational symmetry and gauge invariance
of the original theory become now the internal Spin(d) and global $SU(N)$
symmetry. The physical states become now the $SU(N)$ singlets. We denote the
spatial components of gauge field $A^i_a(t)$ by $x^i_a$ and their
conjugate momenta by $p^i_a$, $[x^i_a,p^j_b]=\delta^{ij}\delta_{ab}$.
The hamiltonian is then [1]

\begin{equation}
H= \frac{1}{2}p^i_a p^i_a + \frac{1}{4}g^2 (f_{abc} x^i_b x^j_c)^2  +H_F,
\end{equation}

\noindent where $H_F=-\frac{i}{2}gf_{abc} \vartheta^{\alpha}_a
x^i_b \Gamma^i_{\alpha \beta} \vartheta^{\beta}_c $ for $D=2,10$
and $\vartheta$ are real spinors obeying $\{
\vartheta^{\alpha}_a,\vartheta^{\beta}_b \}=\delta^{\alpha
\beta}\delta_{ab}$, $\alpha,\beta=1,\ldots,\mathcal{N}$ or $H_F=
igf_{abc} \bar{\vartheta}^{\alpha}_a  x^i_b \Gamma^i_{\alpha
\beta} \vartheta^{\beta}_c $ for $D=4,6$ and $ \vartheta$ are
complex spinors obeying $\{
\bar{\vartheta}^{\alpha}_a,\vartheta^{\beta}_b \}=\delta^{\alpha
\beta}\delta_{ab}$ for
$\alpha,\beta=1,\ldots,\frac{\mathcal{N}}{2}$. The
$\Gamma^i_{\alpha \beta}$  are matrix representation of an $SO(d)$
Clifford algebra $ \{ \Gamma^i,\Gamma^j \}=2\delta^{ij}$.

The growing interest in these models is due to the BFSS ( Banks,
Fischler, Shenker, Susskind ) conjecture [4] where the $N \to
\infty$ limit of Eg.(1) is argued to describes M-theory in the
infinite momentum frame. This stimulated further work on
asymptotic form of the ground state of D=9+1, SU(2), SYMQM [8] and
the analysis of Witten index of (1). The index does not vanish only in
 D=10 where it is equal to 1 [9,10,11,12].
Despite the relevance to M-theory SYMQM have been studied earlier
in different context. The bosonic part of (1) was discovered in
pure Yang-Mills theory in the zero volume limit [2]. Later on it
appeared as a regularization describing the quantum supermembrane
[3]. The detailed study of the hamiltonian (1) shows that in
bosonic sector the potential is confining and there is no
continuous spectrum [6]. If however the supersymmetry is turned on
then  there are bound states in fermion rich sectors as well as
scattering ones [7].

 The only exact solutions of (1) existing in the
literature are for D=1+1, $SU(2)$ [1] and its generalization for
arbitrary $SU(N)$ [5]. Therefore any numerical approach is of interest.

The plan of this paper is the following. In section 2 we briefly
outline the method used to study the models just described and
quote existing results in D=1+1,3+1,9+1 for $SU(2)$ group. In
section 3 and 4 we study general properties in D=1+1 for arbitrary
$SU(N)$ and present the results in D=1+1, $SU(3)$.

\section{Cutoff method}

The cutoff method [13] consists of numerical analysis of the hamiltonian
in the occupation number representation. First we introduce the bosonic and
fermionic creation and annihilation operators ${a^{\dagger}}^i_a$,
$a^i_a$, ${f^{\dagger}}^{\alpha}_a$, $f^{\alpha}_a$ i.e.

\[
a^i_a=\frac{1}{\sqrt{2}}(x^i_a+ip^i_a), \ \ \ \
[a^i_a,{a^{\dagger}}^j_b]=\delta^{ij}\delta_{ab}, \ \ \ \
\{f^{\alpha}_a,{f^{\dagger}}^{\beta}_b\}=\delta^{\alpha
\beta}\delta_{ab}  \footnote{There are several choices of
fermionic $f^{\alpha}_a$, ${f^{\dagger}}^{\alpha}_a$ operators.
Since we do not make any explicit calculations here we refer the
reader to [13] for details.}.
\]

\noindent Next we truncate the Hilbert space to the maximal number
o quanta

\[
n_B= \sum_{i,b} {a^{\dagger}}^i_b a^i_b, \ \ \ \ n_B \leq n_{Bmax},
\]
compute matrix elements of H and diagonalize the resulting
finite matrix. In this way one can analyze the spectrum dependence
on a cutoff ${n_{Bmax}}$. There is a dramatic  difference between  the behavior of the continuous
and discrete spectrum  with cutoff. Namely
\[
E^{n_{Bmax}}_m=E_m+O(e^{-n_{Bmax}})  \ \ - \ \ \hbox{discrete spectrum,} \\
\]
\[
E^{n_{Bmax}}_m=O(\frac{1}{n_{Bmax}})  \ \ - \ \ \hbox{continuous spectrum,}
\]
where m is an index of the energy level $m=1,\ldots,n_{Bmax}+1$. The limit
$n_{Bmax} \longrightarrow \infty $ is called the continuum limit. In
the  case of the discrete spectrum case the energy levels converge rapidly to
the exact eigenvalues of the hamiltonian. This may not be
surprising, however it is interesting to see how fast is the
convergence. For details the reader is referred to [14]. In the
continuous spectrum case things are different. The convergence is
very slow and all the eigenvalues vanish in the infinite cutoff limit.
In the continuum limit the spectrum is continuous and the only way to restore it from cut Fock space
 is to put the following scaling 
[15]

\begin{equation}
m({n_{Bmax}})=const.\sqrt{{n_{Bmax}}} \Longleftrightarrow
E^{{n_{Bmax}}}_{m({n_{Bmax}})}\rightarrow E.
\end{equation}

It was claimed in [15] that this scaling law should work
independently of the theory whenever one can define scattering
states asymptotically. The argument for the above claim is based
on the following fact. The eigenvalues of the momentum operator in
ordinary d=1 quantum mechanics in cut Fock space are zeros of
Hermite polynomials $H_{n_{Bmax}}(x)$ the asymptotic behavior of
which is $\frac{1}{\sqrt{n_{Bmax}}}$ [14,15]. Therefore, once the
momentum operator is defined, its spectrum cutoff dependence
should be $\frac{1}{\sqrt{n_{Bmax}}}$ for large $n_{Bmax}$.

The $E^{n_{Bmax}}_m$ values for fixed $n_{Bmax}$ give the
opportunity to calculate  regularized ($n_{Bmax}$ dependent)
Witten index. If the spectrum of the supersymmetric hamiltonian H
is discrete then the index counts the difference between bosonic
$n^0_b$ and fermionic $n^0_f$ ground states  i.e.

\[
I_W=Tr(-1)^Fe^{-\beta H}=\sum_{m} (-1)^{F(m)}e^{-\beta E_m}=n^0_b-n^0_f,
\]
where F is a fermion number. This quantity is $\beta$ independent.
The cutoff makes it $\beta$ and $n_{Bmax}$ dependent i.e.

\begin{equation}
I^{reg}_W(\beta,n_{Bmax})=\sum_{m=1}^{n_{Bmax}+1} (-1)^{F(m)} e^{-\beta E^{n_{Bmax}}_m}.
\end{equation}

If the spectrum of the hamiltonian H is continuous then the $I_W$
depends on $\beta$ and the difference $n^0_b-n^0_f$ may be
obtained by taking the $\beta \to \infty$ limit. On the other hand the $\beta \to 0$
limit is easier to compute, therefore one introduces the boundary
term $\delta I_W$ using the following trick [9]

\[
\delta I_W=I_W(\infty)-I_W(0)=\int_0^{\infty}d\beta \frac{d}{d\beta}I_W(\beta).
\]

\subsection{D=1+1,3+1,9+1  SU(2) SYMQM }
 In $D=1+1$ case the  hamiltonian $H=\frac{1}{2}p_a p_a+gx_aG_a$, 
 where $G_a$ is the SU(N) generator, is free in a gauge invariant sector.
 There are as many fermion sectors as the grassmann algebra allows i.e. 1 boson sector and $N^2-1$ fermionic
sectors. Since the gauge group is $SU(2)$ we will denote them as
$\mid F=0 \rangle$, $\mid F=1 \rangle$, $\mid F=2 \rangle$, $\mid
F=3 \rangle$. We also have the particle-hole symmetry which
relates sectors $\mid 0 \rangle \leftrightarrow \mid 3 \rangle$
and $\mid 1 \rangle \leftrightarrow \mid 2 \rangle$ hance the
analysis of the first two sectors is sufficient. There is also
supersymmetry which relates sectors $\mid 0 \rangle
\leftrightarrow \mid 1 \rangle$  and $\mid 2 \rangle
\leftrightarrow \mid 3 \rangle$ therefore the whole information about
the spectrum is in fact in the first sector. Supersymmetry does
not communicate between sectors $\mid 1 \rangle$ and $\mid 2
\rangle$ which is exceptional for $SU(2)$. Since the particle-hole
symmetry relates sectors with different fermion number, it is
evident that the regularized Witten index of this model vanishes. It is
however interesting to compute the restricted Witten index which
is defined in first two sectors only and the exact answer is
$\frac{1}{2}$ [16] which was also confirmed numerically.

In D=3+1 dimensions the hamiltonian (1) is not free due to the
quartic potential term.
 There are 6 fermionic sectors. The particle-hole symmetry
relates sectors with the same fermion number hance the eigenstates
from these sectors do not cancel under the sum (3). The analysis
of the index [10] shows that in this case
\[
I_W(\infty)=I_W(0)+\delta I = \frac{1}{4}-\frac{1}{4}=0 \ \ \ \
\hbox{Witten index for D=3+1, SU(2).}
\]
\noindent On the contrary the cutoff analysis  gives the non zero
value [17]. The index converges towards $\frac{1}{4}$ which
is exactly the value of the $I_W(0)$ not $I_W(\infty)$. It seems that the
cutoff method somehow does not contain the boundary term $\delta I$.

This model is the first non-trivial one where the scaling (2) was
confirmed i.e. the spectrum of a free particle $p^2/2$ can be recovered provided eqn. (2) is applied.
 Moreover, in fermion rich sectors both discrete and continuous spectrum is
present which precisely corresponds to conclusions of [7].

The analysis of the supermultiplets is even more interesting. Each
eigenstate is labelled by three quantum numbers: energy E, angular
momentum l and fermion number F. Therefore each state can be
represented by a dot in $R^3$ space. It can be proved [17] that supersymmetry
links these dots in such a way that the emerging geometrical object representing
each supermultiplet is a diamond. This picture very nicely catalogues all the
supermultiplets and it is independent of a gauge group.

In $D=9+1$ dimensions case we only note the astonishing
difficulties that emerge [18]. Since we have the SO(9) symmetry the second order Casimir operator is
\[
J^2=\sum_{i<k}J_{ik}, \ \ \ \ J_{ik}=x^{[i}_a p^{k]}_a
+\frac{1}{2}\psi^{\dagger}_a \Sigma^{ik} \psi_a, \ \ \ \ \Sigma^{ik}=-\frac{i}{4}[\Gamma^i,\Gamma^k].
\]
Normally we would have expect the SO(9) singlet to be the Fock
vacuum $\mid 0 \rangle$. This is not he case here since one can
prove that $J^2\mid 0 \rangle=78 \mid 0 \rangle$ [18]. The empty state is not invariant under rotations! This is a
surprising fact and it means that the SO(9) singlet is somewhere
else. Where is it? The model has 24 fermionic sectors and it was found that the
singlet happens to be just in the central F=12 sector.

\section{The general properties ot the D=1+1, SU(N) SYMQM }

 Since the eigenstates in SYMQM are the gauge singlets therefore it is
reasonable to ask about the convenient $SU(N)$ invariant basis.
It is evident that states belonging to such basis have to be of the form

\begin{equation}
T_{bc\ldots de\ldots}a^{\dagger}_b a^{\dagger}_c \ldots
f^{\dagger}_b f^{\dagger}_c \ldots \mid 0 \rangle,
\end{equation}
where $T_{bc\ldots de\ldots}$ is some SU(N) invariant tensor made
out of structure tensors $f_{abc},d_{abc},\delta_{ab}$. We now proceed to chose linearly
independent states from (4).

\subsection{Birdtracs}

In order to deal with the variety of all possible tensor contractions
we introduce the diagrammatic approach (figure 1).

\begin{figure}[h]
\centering \leavevmode
\includegraphics[width=0.6\textwidth]{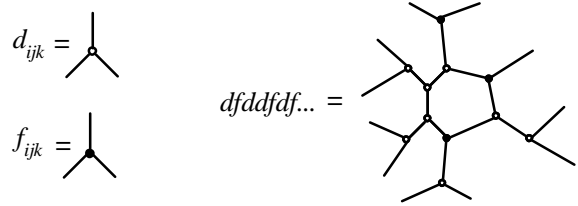}
\caption{Diagrammatic notation of invariant tensors.}
\end{figure}

Each leg corresponds to one index and summing over any two indices
is simply gluing appropriate legs. Structure tensors $f_{ijk}$,
$d_{ijk}$ are represented by vertices and $\delta_{ij}$ is a line.
Any tensor may now be represented by a graph. Such diagrammatic
approach has already been introduced long time ago by
Cvitanovi\v{c} [19]. In,
general one can construct loop tensor which by definition is a
tensor that diagrammatically looks like a loop however it can be
proved [20] that any such loop can be expressed in terms of
forests i.e. products of tree tensors  ( figure 2)

\begin{figure}[h]
\centering
\includegraphics[width=0.9\textwidth]{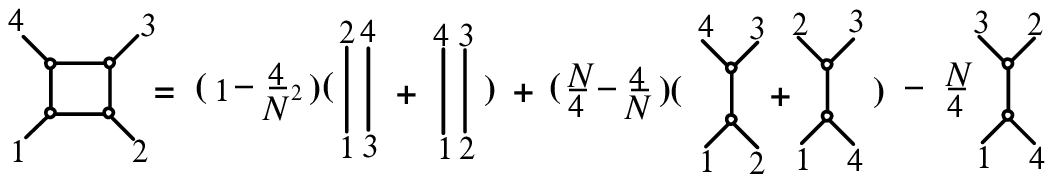}
\caption{An example of loop reduction for a square made out of
$d_{ijk}$ tensors.}
\end{figure}
\noindent Therefore we are left with tree tensors only. These however can be
easily expressed in terms of trace tensors  $Tr(T_a T_b \ldots)$
where $T_a$  are  $SU(N)$ generators in fundamental
representation. With the use of the following matrices
$A^{\dagger}=a^{\dagger}_bT_b$, $F^{\dagger}=f^{\dagger}_bT_b$ any
gauge invariant state can be obtained by acting with an appropriate linear combination of
products of trace operators

\[
Tr({A^{\dagger}}^{i_1}F^{\dagger}{A^{\dagger}}^{i_2}F^{\dagger}\ldots
{A^{\dagger}}^{i_k}F^{\dagger}),
\]
on Fock vacuum $\mid 0 \rangle$.
Due to the grassmann algebra the number of F matrices under the
trace cannot be grater then $N^2-1$ i.e. $k\leq N^2-1$. Moreover
the Cayley-Hamilton theorem for A matrices gives $i_k \leq N$. The
remaining set of states is still linearly dependent and the
further analysis requires separate study of each $SU(N)$. The
basis states in F=0 sector are of the form

\[
\mid i_2,i_3,\ldots,i_N \rangle=Tr^{i_2}({A^{\dagger}}^2)
Tr^{i_3}({A^{\dagger}}^3)\ldots
Tr^{i_N}({A^{\dagger}}^N)\mid 0 \rangle.
\]

\noindent We see that there are as many states with given number
of quanta $n_B$ as there are natural solutions of the equation
$2i_2+3i_3+\ldots+Ni_N=n_B$. For $U(N)$ this would be exactly
$p(n_B)$ - the partition number of $n_B$. For $SU(N)$ this is a
little less then $p(n_B)$ however it still grows exponentially with
$n_B$.

 In order to solve the model in bosonic sector one has to
compute the following scalar product

\[
\textit{N}^{i_2 \ldots i_N}_{j_2 \ldots j_N}=\langle i_2 \ldots i_N \mid
j_2 \ldots j_N \rangle,
\]

\noindent which in principle is a tedious, but not impossible, task .

Let us discuss, the "bilinear" basis which by definition is the following restricted $SU(N)$ basis

\begin{equation}
\mid  2n \rangle=(A^{\dagger}A^{\dagger})^n\mid 0 \rangle, \ \ \ \ (A^{\dagger}A^{\dagger})=a_i^{\dagger}a_i^{\dagger}
\end{equation}

\noindent which was introduced in [2] in D=3+1 case. In this
basis the non zero hamiltonian matrix elements are easy to derive.
First we write the commutation relations

\begin{equation}
[(AA),(A^{\dagger}A^{\dagger})^n]=4n(A^{\dagger}A^{\dagger})^{n-1}(A^{\dagger}A)+4n(n-1+\frac{N^2-1}{2})(A^{\dagger}A^{\dagger})^{n-1},
\end{equation}

\begin{equation}
[(AA^{\dagger}),(A^{\dagger}A^{\dagger})^n]=2n(A^{\dagger}A^{\dagger})^n.
\end{equation}
Using (6) we obtain norms for $\mid  2n \rangle$ i.e.

\[
c^2_{2n}:=\langle 2n \mid 2n \rangle=
4n(n-1+\frac{N^2-1}{2})c^2_{n-1}, \  \  \  \  c_{2n}=\sqrt{
\prod_{k=1}^n 4k(k-1+\frac{N^2-1}{2})}, \ \ \ \ c_0=1.
\]
In the orthonormalized basis $ \tilde{\mid  2n
\rangle}=\frac{1}{c_{2n}}\mid  2n \rangle$ the non vanishing
matrix elements of the hamiltonian

\[
H=\frac{1}{2}p_ap_a=-\frac{1}{4}((A^{\dagger}A^{\dagger})+(AA)-2(A^{\dagger}A)-(N^2-1)),
\]
are

\[
\tilde{\langle 2n \mid} H \tilde{\mid 2n \rangle}=n+\frac{N^2-1}{4}
\]
and
\[
\tilde{\langle 2n+2 \mid} H \tilde{\mid 2n \rangle}
=\tilde{\langle 2n \mid} H \tilde{\mid 2n+2
\rangle}=-\frac{1}{2}\sqrt{(n+1)(n+N^2-1)}.
\]
Therefore it is straightforward to proceed with the cutoff
analysis (figure 3). We see that thee is no quantitative difference between $SU(2)$ and
eg. $SU(100)$ case. This is not what we have expected and it means
that the restricted basis (5) simplifies too much.

\begin{figure}[h]
\centering
\includegraphics[width=0.9\textwidth]{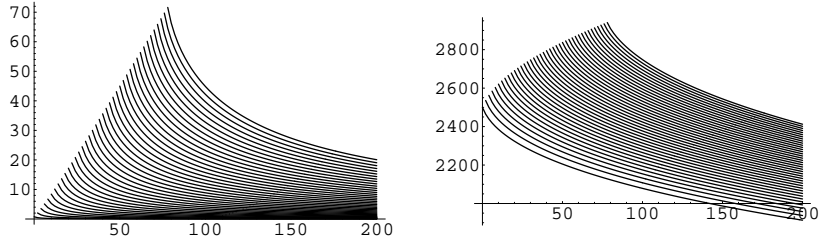}
\caption{The cutoff dependence of spectrum for $SU(2)$ and
$SU(100)$ in "bilinear" basis.}
\end{figure}

\section{ D=1+1, SU(3) SYMQM }

Here we present the calculations of Hamiltonian matrix
elements in a complete basis in bosonic sector. The basis vectors and the scalar products, we are interested,
in  are

\begin{equation}
\mid i,j \rangle =
(A^{\dagger}A^{\dagger})^i (A^{\dagger}A^{\dagger}A^{\dagger})^j\mid
0 \rangle, \ \ \ \ \textit{N}^{i \ j}_{i' \ j'}= \langle i,j \mid i',j'
\rangle.
\end{equation}
\[
(A^{\dagger}A^{\dagger}A^{\dagger})=d_{ijk}a_i^{\dagger}a_j^{\dagger}a_k^{\dagger}
\]
The only non vanishing elements of $S^{i \ j}_{i' \ j'}$ are the
ones obeying the constraint $2i+3j=2i'+3j'$. Therefore it is
convenient to work with the following symbol

\[
W^k_{i \ j} = \langle i, j \mid  (AAA)^{2k}(A^{\dagger}A^{\dagger})^{3k} \mid i,j
\rangle,
\]

\noindent which has the advantage of reproducing all non
vanishing $\textit{N}^{i \ j}_{i' \ j'}$'s. It is tedious but possible to
obtain formulas and recurrence equations for $W^k_{ij}$. We shall
omit the lengthy derivation and only give the results.

First we solve the recurrences for $W^k_{00}$ and $W^k_{i0}$. We have

\begin{equation}
W^k_{00}=96k(2k-1)(9k^2-1)(9k^2-4)W^{k-1}_{00}, \ \ \ \ 
W^k_{i0}=4(3k+i)(3k+i+3)W^{k}_{i-1  \  0},  \ \ \  \ W^0_{00}=1.
\end{equation}
therefore (9) gives an exact formula for $W^k_{i0}$. The
$W^k_{0j}$ term is computed from the following recurrence
\[
W^k_{0 j}=\alpha_{jk} W^{k-1}_{0 j}+\beta_{jk}W^{k}_{0
j-2}+\gamma_{jk} W^{k+1}_{0 j-4},
\]
where
\[
 \alpha_{jk}=48(2k+j)(2k+j-1)(3k-1)(3k-2)(3k+3j+2)(3k+3j+1),
\]
\[
\beta_{jk}=72(2k+j)(2k+j-1)j(j-1)(9k^2+9kj-2),
\]
\[
\gamma_{jk}=27(2k+j)(2k+j-1)j(j-1)(j-2)(j-3).
\]
This recurrence stops on  $W^k_{0j}$ given by (9).  The general
term $W^k_{ij}$ is now computed from yet another recurrence
\[
W^k_{ij}=4(i+3k)(i+3k+3j+3)W^k_{i-1 j}+3j(j-1)W^{k+1}_{i-1 j-2},
\]
which stops on $W^k_{0j}$ and $W^k_{i0}$.
The whole norm matrix (8) can now be computed. It should be noted that the elements of the N matrix were obtained independently by computing the scalar products $ \langle i,j \mid
i',j' \rangle$ with use of the program written in Mathematica
[13]. In this way all the recurrences presented here were confirmed up to $n_B=12$ i.e. for $(i,j)$ such that $2i+3j \le 12$. This matrix is in fact the Gram matrix which indicates that we still have to
orthogonalize the basis. We will not do so however. In order to
represent the hamiltonian $H$ in orthogonal basis we follow [21]. It
is sufficient to calculate the Gram matrix $G$ and proceed with
the following similarity transformation

\[
H_{ort}=G^{-\frac{1}{2}}H G^{-\frac{1}{2}}.
\]
The results of the cutoff analysis are presented in figure 4.

\begin{figure}[h]
\centering
\includegraphics[width=0.6\textwidth]{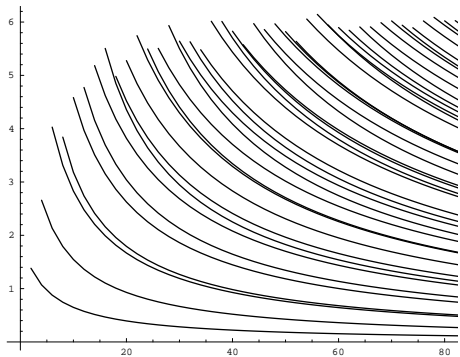}
\caption{The cutoff dependence of spectrum in $D=1+1$, $SU(3)$,
$F=0$.}
\end{figure}

It is clear that the spectrum seems to be far more complicated
than in $SU(2)$ case. The lines in figure 4 are divided into
groups where they converge together.  This can be understood in
the following way. In $SU(3)$ we have two Casimir operators
$T_aT_a$ and $d_{abc}T_aT_bT_c$
 where $T_a$'s are $SU(3)$ generators.
In cut Fock space the second one does not commute with the
hamiltonian therefore the cutoff $n_B$ breaks the $SU(3)$
symmetry. In $n_B \to \infty$ limit the symmetry should be
restored which corresponds to grouping of the lines in figure 4.

\section{Summary}
SYMQM models reveal variety of application in several areas of
physics ( Yang-Mills theories, supersymmetry, strings) hance their
detailed analysis is of interest. Although they are rich in
symmetries ( $SU(N)$ , SO(d), supersymmetry ) the exact solutions
are missing in the literature forcing one to apply numerical
methods. The cutoff method presented here is  working surprisingly
well, however to get any of results of sections [2,3,4] one had to
employ a lot of theoretical work which in some cases gave exact
results (e.g. the structure of supermultiplets). The analysis of
D=1+1 SYMQM for arbitrary $SU(N)$ is very encouraging and gives a
hope to proceed with the $N \to \infty$ limit.

\section{Acknowledgments}
I thank the organizers of the XLV'th Cracow School of
Theoretical Physics and the organizers of the XXIIIrd International
Symposium on Lattice Field Theory for their kind invitation and support. I also thank J. Wosiek for discussions.
This work was supported by the Polish Committee for Scientific Research
under grant no. PB 1P03B 02427 (2004-2007 ).

\end{document}